\documentclass[aps,prl,superscriptaddress,twocolumn,showpacs]{revtex4}

\usepackage{graphicx,color,ulem}

\newcommand{\YRS}{YbRh$_2$Si$_2$}

\begin{document}


\title{Evolution of the Kondo state  of YbRh$_2$Si$_2$ probed by high field ESR}

\author{U. Schaufu\ss}
\affiliation{IFW Dresden, Institute for Solid State Research, P.O.
Box 270116, D-01171 Dresden, Germany}

\author{V. Kataev}
\email{V.Kataev@ifw-dresden.de} \affiliation{IFW Dresden, Institute
for Solid State Research, P.O. Box  270116, D-01171 Dresden, Germany}

\author{A. A. Zvyagin}
\affiliation{IFW Dresden, Institute for Solid State Research, P.O. Box  270116, D-01171 Dresden, Germany} \affiliation{B.I. Verkin Institute for Low
Temperature Physics and Engineering, NAS,  Kharkov, 61103, Ukraine}

\author{B. B\"uchner}
\affiliation{IFW Dresden, Institute for Solid State Research, P.O.
Box  270116, D-01171 Dresden, Germany}

\author{J.~Sichelschmidt}
\email{Sichelschmidt@cpfs.mpg.de}
\affiliation{Max Planck Institute for Chemical Physics of Solids, 01187 Dresden, Germany}

\author{J. Wykhoff}
\affiliation{Max Planck Institute for Chemical Physics of Solids, 01187 Dresden, Germany}

\author{C. Krellner}
\affiliation{Max Planck Institute for Chemical Physics of Solids, 01187 Dresden, Germany}

\author{C. Geibel}
\affiliation{Max Planck Institute for Chemical Physics of Solids, 01187 Dresden, Germany}

\author{F. Steglich}
\affiliation{Max Planck Institute for Chemical Physics of Solids, 01187 Dresden, Germany}

\date{\today}

\begin{abstract}
An electron spin resonance (ESR) study of the heavy fermion compound \YRS\ for fields up to $\sim 8$\,T reveals a strongly anisotropic signal below
the single ion Kondo temperature $T_K\sim 25$\,K. A remarkable similarity between the $T$-dependence of the ESR parameters and that of the specific
heat and the $^{29}$Si nuclear magnetic resonance data gives evidence that the ESR response is given by heavy fermions which are formed below $T_K$
and that ESR properties are determined by their field dependent mass and lifetime. The signal anisotropy, otherwise typical for Yb$^{3+}$ ions,
suggests that, owing to a strong hybridization with conduction electrons at $T<T_K$, the magnetic anisotropy of the 4$f$ states is absorbed in the
ESR of heavy quasiparticles. Tuning the Kondo effect on the 4$f$ states with magnetic fields $\sim 2 - 8$\,T and temperature $2-25$\,K yields a
gradual change of the ESR $g$-factor and linewidth which reflects the evolution of the Kondo state in this Kondo lattice system.
\end{abstract}


\maketitle

\begin{figure}
\begin{center}
\includegraphics[width=\columnwidth,clip]{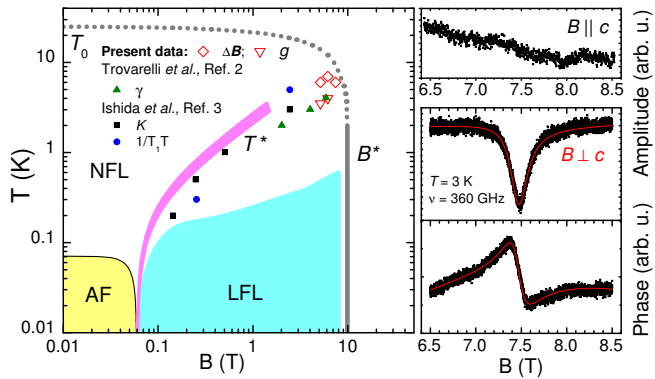}
\caption{(color online) Left: Schematic phase diagram of \YRS\ for $B\perp c$ from Ref.~\onlinecite{Gegenwart08}. Dash
and solid gray lines delineate the region of heavy QP behavior confined to the region $T<T_0$ and $B<B^*$
\cite{Trovarelli00,Tokiwa05,Gegenwart06}. The LFL region denotes the $T-B$ domain where the resistivity follows LFL
behavior $\rho\propto T^2$. The broad (red online) line $T^*$ depicts the position of crossover in the isothermal Hall
resistance, magnetostriction, magnetization, and longitudinal resistivity. Closed symbols depict $T(B)$-crossover
temperatures below which specific heat ($\gamma$)\cite{Trovarelli00} and $^{29}$Si-NMR quantities ($K$ and $1/(T_1T)$)
\cite{Ishida02} become $T$ independent. Open symbols display crossover temperatures from HF-ESR $g(T,B)$- and $\Delta
B(T,B)$ dependences (see text). Right:  A representative HF-ESR signal for $B\perp c$-axis. Absorptive (amplitude) and
dispersive (phase shift) parts of the ESR signal (noisy curves) are simultaneously fitted to a Lorentzian function
(solid lines). No signal can be observed for $B\parallel c$-axis suggesting a strong $g$-factor
anisotropy.}\label{diag_spectra}
\end{center}
\end{figure}

\begin{figure}
\begin{center}
\includegraphics[width=0.75\columnwidth,clip]{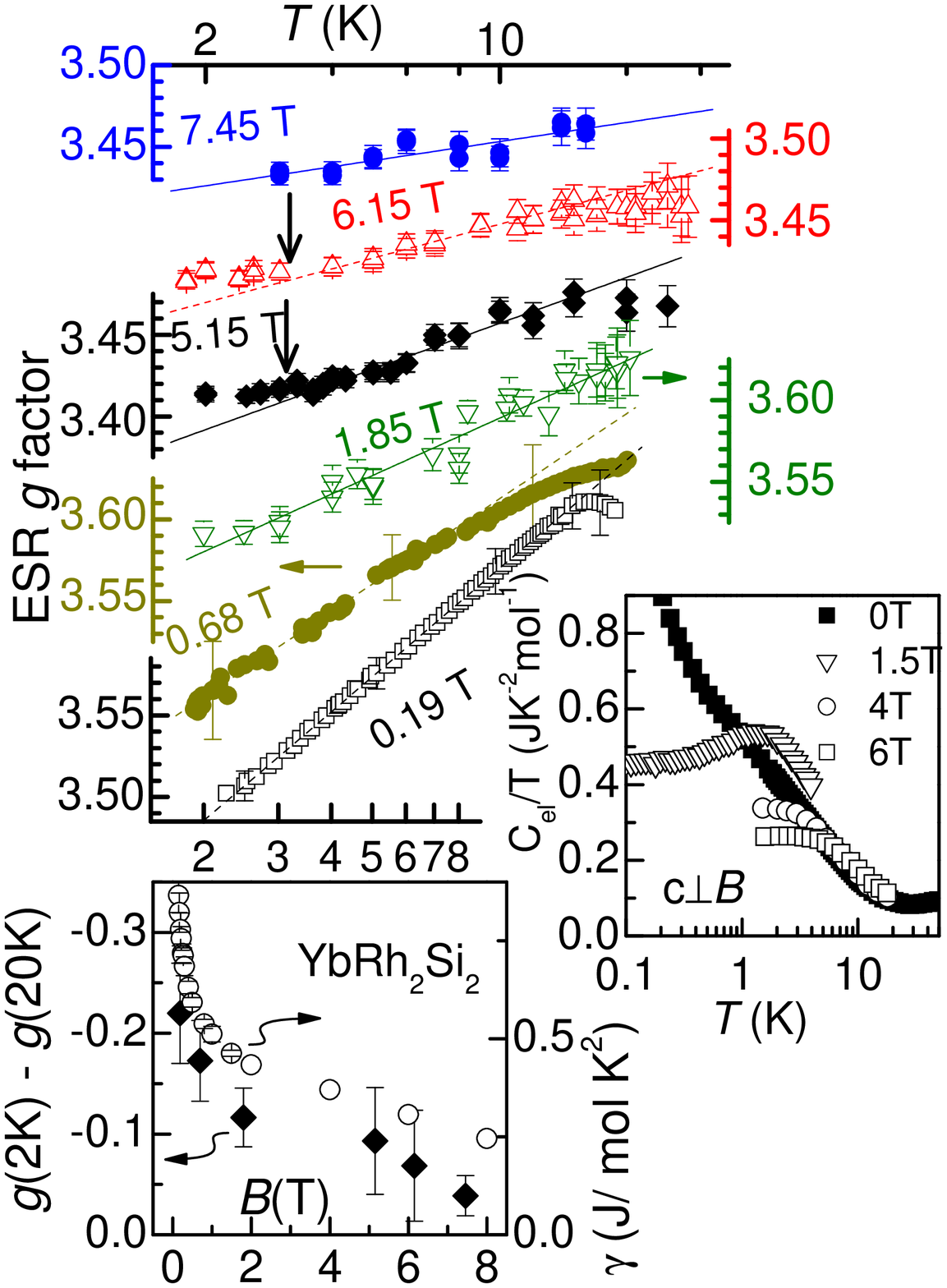}
\caption{ Upper panel: $g(T)$-dependence for $B\perp c$ at $\nu = 9, 34, 93, 249, 297$  and 360\,GHz (from bottom to
top) which correspond to resonance fields as indicated. Arrows indicate a crossover from a $-\ln T$ behavior to the $g
=$ const regime. Inset: $T$-dependence of the Sommerfeld coefficient $\gamma=C_{el}(T)/T$ at different magnetic fields
(Ref.~\onlinecite{Gegenwart06}).  Lower panel: Field dependence of low temperature $\gamma$
(Ref.~\onlinecite{Gegenwart06}) and $g$-shift between 20\,K and 2\,K. For details see text.} \label{gfactor}
\end{center}
\end{figure}

\begin{figure}
\begin{center}
\includegraphics[width=0.75\columnwidth,clip]{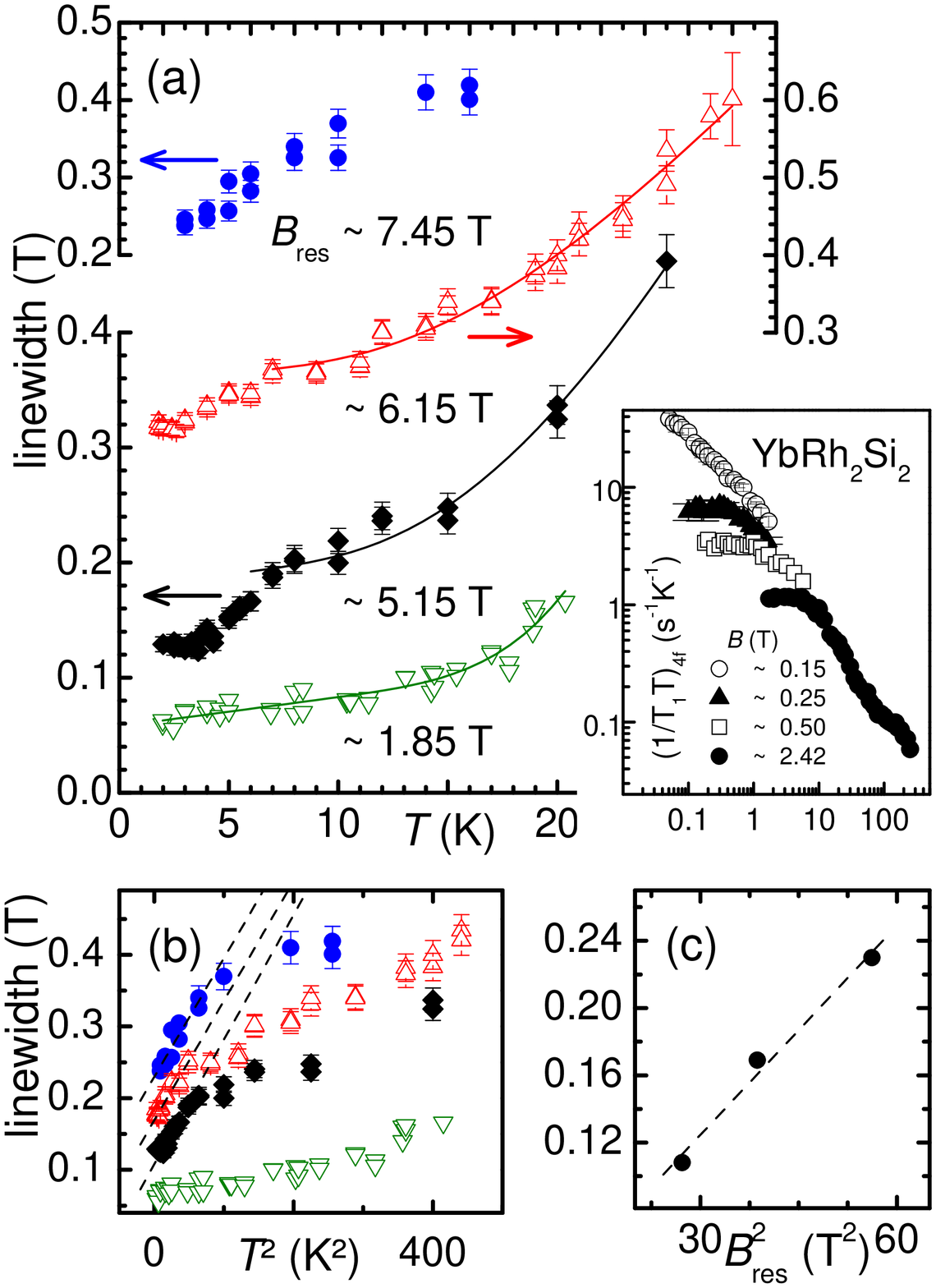}
\caption{(a) - $T$-dependence of the linewidth $\Delta B$ for $B\perp c$ at $\nu = 93$ (diamonds) 249 (squares), 297 (circles) and 360\,GHz
(triangles). Solid lines are fits (see the text). The inset shows the $T$-dependence of $^{29}$Si relaxation rate $1/(T_1T)$ from
Ref.~\onlinecite{Ishida02}. (b) - $\Delta B$ vs. $T^2$ representation of the data. Dashed lines are fits of the low-$T$ data points to the
$T^2$-dependence. (c) - residual width at $T = 0$ from the $T^2$-fit plotted vs. $B_{\rm res}^2$. }
  \label{width}
\end{center}
\end{figure}

Strong electron-electron (EE) interactions in metals yield a fascinating variety of novel and often interrelated quantum phenomena, such as quantum
phase transitions, breakdown of the Landau Fermi-liquid (LFL) state, unconventional superconductivity, etc. (for an overview see, e.g.,
\cite{Gegenwart08}). In intermetallic compounds where $4f(5f)$ magnetic ions (e.g. Yb, Ce, U { \it etc.}) build up a regular Kondo lattice, strong EE
correlations are established by the coupling of local $f$-magnetic moments with the conduction electrons (CE). As a consequence, a large effective
mass enhancement of the quasiparticles (QP) hallmarks the properties of paramagnetic heavy fermion metals. A competing interaction, the so-called
RKKY-interaction between the local $f$-states
via the sea of CE, favors a magnetically ordered ground state.\\
\indent An important realization of a system where the delicate balance between Kondo and RKKY interactions can be investigated is the intermetallic
compound \YRS\ where antiferromagnetic order, quantum criticality, heavy fermion- and non-LFL (NFL) behavior can be tuned by a magnetic field $B$ and
temperature $T$ \cite{Trovarelli00,Ishida02,Tokiwa05,Gegenwart06,Gegenwart06b} (Fig.~\ref{diag_spectra}).
In the parameter domain where these remarkable electronic crossovers take place a strong hybridization of 4$f$ electrons with CE significantly
broadens the otherwise atomically sharp $f$-states. That is why the observation of a narrow electron spin resonance (ESR) signal in the Kondo state
of \YRS\  was very surprising \cite{Sichelschmidt03}. While the reported pronounced anisotropy of the signal is indeed in accordance with an ESR of
localized Yb$^{3+}$ $4f$ moments, a non-local picture is suggested by the observation of this signal down to the lowest accessible temperatures of
0.69\,K \cite{Wykhoff07} where the single ion Kondo effect is expected to screen the magnetic moments. On the other hand the conduction electron ESR
seems also unlikely because in this compound comprising heavy metal elements the spin-orbit (SO) coupling
drastically shortens the electron spin lifetime \cite{Monod79}.\\
\indent To unravel a controversial nature of this resonance response we have studied ESR of a high quality single crystal of \YRS\ in a broad
magnetic field region varying between the low-field/low-frequency (LF) ($\sim 0.2$\,T/ $ \nu\sim 9$\,GHz) and the high-field/high-frequency (HF)
($\sim 8$\,T/ $ \nu\sim 360$\,GHz) regimes. In the LF limit \YRS\ is in the NFL state whereas in the HF regime it is in a LFL state close to the
breakdown of the heavy-fermion behavior which is confined to temperatures and fields $T < T_0\approx 25$\,K and $B < B^*\approx 10$\,T
\cite{Tokiwa05}. Here $T_0$ denotes the characteristic spin fluctuation temperature (which corresponds to the single ion Kondo temperature $T_K$)
\cite{Trovarelli00} and $B^*$ is a characteristic field which separates more itinerant from more localized behavior of the 4$f$ states
\cite{Trovarelli00,Tokiwa05,Gegenwart06}. In this parameter domain electronic specific heat (related to the density of states of QPs)
\cite{Trovarelli00,Gegenwart06} and nuclear magnetic resonance (NMR) data (related to the spindynamics of QPs) \cite{Ishida02} indicate the
development of heavy fermion behavior with a field dependent crossover to clear LFL properties at low $T$ \cite{Gegenwart06}. We find that this
development and crossover behavior is reflected also in the temperature- and field dependence of the ESR response. This similarity gives evidence
that ESR in \YRS\ is given essentially by the resonance of heavy fermions providing thus \textit{direct} experimental access to the dynamics of heavy
quasiparticles in the Kondo state.\\
\indent LF-ESR was measured with a standard high-sensitive cavity based reflection technique \cite{Sichelschmidt03}. HF-ESR was carried out in a
reflection geometry with a Millimeterwave Vector Network Analyzer (AB Millimetre) at $\nu = 93-360$\,GHz in a field range 0 - 14\,T.
We used a very sensitive induction mode scheme that detects the change of the microwave's polarization at the resonance \cite{Fuchs99}.\\
\indent The strong anisotropy of the ESR response corresponds to the strong magnetic anisotropy of \YRS\ \cite{Sichelschmidt07, Trovarelli00} and is
nicely reflected in the HF-ESR data: For $B\parallel\,c$-axis no ESR response has been found, whereas a well defined resonance line with a $g$-factor
$\sim 3.5$ has been observed for $B\perp\,c$ (Fig.~\ref{diag_spectra}, right). Note that our HF ESR setup allows to measure both the amplitude and
the phase shift of the signal. This enables a separation of the absorption and dispersion part of the complex resonance response of a metallic
sample. Simultaneous fitting of the amplitude and the phase shift at the resonance (Fig.~\ref{diag_spectra}, right) provides an accurate
determination of the resonance field $B_{\rm res}$ and the linewidth $\Delta B$.\\
\indent The $T$-dependences of the $g$-factor $g = h\nu/\mu_B B_{\rm res}$ and $\Delta B$  at selected excitation
frequencies are summarized in Figs.~\ref{gfactor} and \ref{width}. For large $B_{res}$ the data appreciably scatter
owing to a moderate signal-to-noise ratio, caused by a considerable broadening of the signal with increasing the
temperature and magnetic field strength, as well as the reduction of the microwave penetration depth with increasing
$\nu$. However, two central results can be deduced: (i) the $g$-factor data sets show a decreasing $T$ variation with
increasing magnetic field, and (ii) one can identify distinct regimes with different $T$-dependences of the $g$-factor
and $\Delta B$ in the data sets at $\nu =$ 249, 297, and 360\,GHz ($B_{res}\approx$ 5.15, 6.15, and 7.45\,T). At a
fixed field (or frequency), the $g$-factor (Fig.~\ref{gfactor}) increases approximately as $\ln(T)$ at high $T$s for
all fields. However, $g(T)$ shows a saturation tendency below $\sim 4 - 5$\,K for the data at $\nu = 249$ and 297\,GHz.
At these frequencies the $\Delta B(T)$-data also show an anomaly, namely a broad hump at somewhat higher $T\sim 7 -
8$\,K (Fig.~\ref{width}). This crossover is more emphasized in the
plot $\Delta B$ vs. $T^2$ shown on the same Figure (see below).\\
\indent As a function of temperature and magnetic field the thermodynamic quantities, such as the electronic specific
heat $C_{el} \propto N(E_F)T$ and the Pauli susceptibility $\chi_p \propto N(E_F)$, both being a measure of the density
of states  $ N(E_F)$ of QPs at the Fermi level $E_F$ and thus their effective mass $m^*$, experience notable
modifications. In the $T-B$ range of the HF-ESR data $\chi_p$ probed by the $^{29}$Si Knight shift $K \propto \chi_p$
and the Sommerfeld coefficient $\gamma = C_{el}/T$ change their $T$-dependences from  $-\ln T$ at higher $T$ to
$\approx $ const below $T \sim 2-4$\,K (Fig.~\ref{gfactor}, inset, and
Ref.~\onlinecite{Trovarelli00,Ishida02,Gegenwart06}). This may be interpreted as the development of the heavy fermion
QPs at higher $T$ followed by the establishment of the coherent LFL state at lower $T$. Remarkably, these changes occur
at approximately the same temperatures and fields where the $\ln T$ dependence of the $g$-factor levels off ($T\sim
2-4$\,K; see arrows in Fig.~\ref{gfactor} and symbols in Fig.~\ref{diag_spectra}, left). A close correspondence between
the increase of $\gamma$ which reflects a strong enhancement of $m^*$ and the $g$-shift at low $T$ can be seen in
Fig.~\ref{gfactor}, lower panel. Moreover, at low $T$ there is an apparent correspondence between the changes of the
$T$-behavior of the dynamic quantities, namely of the longitudinal $^{29}$Si spin relaxation rate $1/T_1$ divided by
$T$, which is proportional to the momentum $q$ averaged dynamical electron spin susceptibility
$\chi^{\prime\prime}(q,\omega)$, and the ESR linewidth $\Delta B(T)$ (Fig.~\ref{width} and Ref.~\onlinecite{Ishida02}).
The crossover temperatures from specific heat, NMR and present HF-ESR data are summarized in the phase diagram in
Fig.~\ref{diag_spectra}. The ESR points obviously fall into a common crossover line. Such remarkable similarities
between the characteristic changes of $\gamma$ and NMR quantities on the one hand, and the HF-ESR data on the other
hand strongly suggest that ESR in \YRS\ is given by the response of the heavy electrons to microwaves and as such it
reflects a crossover between the different electronic regimes. Note that the diffusion time $\tau_D$ of the heavy
electrons in \YRS\ is long enough \cite{Abrahams08} for the ESR lineshape to be Lorentzian, which is a limiting case of
a "Dysonian" at $\tau_D \rightarrow \infty$.

\indent In the following we discuss plausible reasons making such a heavy electron spin resonance experimentally
observable and its relation to the Kondo state of \YRS. Generally, ESR is given by the frequency dependent uniform
($q=0$) transverse electron spin susceptibility. From its pole one obtains the resonance frequency $\nu= g\mu_BB_{\rm
res}/h$ and the linewidth $\Delta B \propto 1/T^*_2$. Here $T^*_2$ is the electron spin-spin dephasing time which in
metals is equal to the spin-lattice relaxation time $T_1$ (see, e.g. Ref.~\onlinecite{Freedman75}). For the spin
resonance of conduction electrons (CESR) $\nu$ and $\Delta B$ depend on the details of the band structure and the SO
interaction \cite{Monod79}. Owing to the large magnitude of the latter in metals containing heavy elements the CESR
signal becomes unobservable \cite{Monod79}. Magnetic resonance of the {\it localized} states in a metallic host can be
observed much easier even in materials with strong SO interaction. This is because the 4$f$-electrons being buried
deeply beneath the outer electron shells of the rare-earth ion retain the sharp atomic like character of the energy
states and generally mix little with the CE. Also in Kondo systems at temperatures $T>T_K$ were the local moments are
still well defined their resonance can be observed experimentally, as, e.g., in a dilute Au:Yb alloy with $T_K \sim
10$\,$\mu$K \cite{Baberschke80}. However, at $T \lesssim T_K$ the signal is expected to disappear owing to the Kondo
screening of the local moments by CE \cite{Sichelschmidt03,Abrahams08}.
A single ion Kondo scenario fails to explain the occurrence of a sharp ESR mode in terms of the resonance of
well-localized Yb$^{3+}$ moments \cite{ZM}. It is similarly unreasonable to discuss the ESR signal considering only
conduction electrons, as if \YRS\ were an uncorrelated metal. In this scenario a strong SO scattering due to the
presence of heavy elements would smear out the resonance \cite{comment_bottleneck}. In the Kondo state a coherent
many-body interaction between itinerant electrons and local moments gives rise to QPs with a strongly enhanced
effective mass, i.e heavy fermions. One can conjecture that the observed ESR response of \YRS\ is a novel type of the
magnetic resonance excitation given by heavy electrons. Owing to a strong hybridization effect the heavy QPs may
inherit to a significant extent the anisotropic properties of the $f$-electrons, specifically a typical strong
anisotropy of the $g$-factor. In this situation the uniform static spin susceptibility of heavy fermions, $\chi_p$,
contributes to the shift of the $g$-factor. This should yield a $T$-dependence similar to that of the electronic
specific heat and the NMR Knight shift which is indeed experimentally observed in \YRS\ (Fig.~\ref{gfactor}). Comparing
with the static susceptibility data \cite{Gegenwart06b} the $g$-shift shows a crossover to the heavy fermion LFL state
even more clearly suggesting that ESR addresses directly the QPs` spin susceptibility. In fact, the field dependence of
the $g$-factor's temperature variation, Fig.~\ref{gfactor}, reflects what qualitatively would be expected for the
development of the heavy fermion state in \YRS\ due to the Kondo effect: with decreasing temperature and/or decreasing
magnetic field the Kondo interaction between 4$f$ and conduction electron spins enhances giving rise to heavy fermions
and the $g$-factor deviates from the values of the completely localized 4$f$ spin. Remarkably, the ESR singal in \YRS\
becomes observable for \textit{all} magnetic fields at temperatures comparable or
smaller than  $T_0 \approx 25$\,K further suggesting that its occurrence is related to the formation of heavy fermions.\\
\indent In Fig.~\ref{width} the $T$-dependence of the linewidth was parameterized as $\Delta B(T) = a +bT +
c/(\exp(\Delta/T) -1 )$. Here $a$ depicts a residual width, $bT$ stands for the relaxation broadening via electronic
degrees of freedom, and the last term has been assigned in Ref.~\onlinecite{Sichelschmidt03} to a relaxation via an
excited magnetic state of Yb$^{3+}$ at an energy $\Delta$ above the ground state. However, the fit of the HF-ESR data
requires a significant reduction  of $\Delta$ from $\sim 115$\,K (Ref.~\onlinecite{Sichelschmidt03}) to $\sim 50 -
60$\,K.
Moreover, from neutron scattering results this excitation energy $\Delta$ is much higher ($\Delta \sim 200$\,K) \cite{Stockert06}. Therefore, the
exponential term in $\Delta B(T)$ is not related to the relaxation via excited states. Instead, it seems that the strong broadening of the ESR
response mimics the breakdown of the heavy fermion state approaching the $T_0/B^*$
crossover line (Fig.~\ref{diag_spectra}).\\
\indent For $B_{res}\sim$  5.15 and 6.15\,T $\Delta B(T)\propto 1/T^*_2(T)$ deviates appreciably from the fit at $T <
7$\,K and turns to a $T^2$-dependence (Fig.~\ref{width}, lower panel) suggesting the occurrence of a new spin
relaxation regime for heavy fermions. We recall that this change is concomitant with characteristic crossovers in the
ESR $g(T)$-dependence (Fig.~\ref{gfactor}, arrows) as well as in the specific heat ($\gamma$) and NMR ($K, 1/(T_1T))$
measurables pointing at the common origin of these features. In particular, the saturation of $\gamma, K$ and
$1/(T_1T)$ at low temperatures has been associated with the establishment of properties of the LFL state
\cite{Trovarelli00,Ishida02,Gegenwart06}. Since classically the
 spin relaxation of QPs is proportional to  their momentum relaxation, a $T^2$ law for $\Delta B(T)$ could be related to the
QP - QP scattering in the LFL state. However, in the resistivity the $T^2$ regime occurs at much lower temperatures \cite{Gegenwart06}
(Fig.~\ref{diag_spectra}) implying a more complex relationship between the spin- and momentum relaxation of heavy electrons in a correlated
quantum metal compared to its classical counterpart.\\
\indent The idea that many-body effects due to strong EE interactions may qualitatively change the ESR response in
correlated metals has been discussed already for quite a time. For instance, the occurrence of a narrow collective ESR
mode with a considerable shift of the $g$-factor has been predicted in Ref. \onlinecite{Freedman75}. Such a sharp mode
is expected for an arbitrary sign of the EE interaction, though owing to the enhanced spin susceptibility the
ferromagnetic (FM) case is easier to observe experimentally. Striking examples for such a situation are the CESR
signals in Pd \cite{Monod78} and TiBe$_2$ \cite{Shaltiel87}. For Kondo lattice systems there is strong experimental
evidence that FM correlations between the Kondo ions are essential for narrow observable ESR signals \cite{Krellner08}.
For \YRS\ in the $T-B$ parameter domain studied in the present work FM fluctuations have been found by $^{29}$Si-NMR
\cite{Ishida02}. Very recently motivated by ESR experiments on \YRS\ two theoretical models of ESR in Kondo lattice
systems with anisotropic (Ref.~\onlinecite{Zvyagin08}) and isotropic (Ref.~\onlinecite{Abrahams08}) magnetic EE
interactions have been proposed. Though using different approaches both theories predict in particular  for the LFL
phase with FM interactions a sharp ESR line only slightly shifted from the position expected for the local 4$f$
resonance. The narrowing takes place by a factor of the mass enhancement $m/m^*$ \cite{Zvyagin08,Abrahams08} or is due
to the action of only the anisotropic part of EE interaction \cite{Zvyagin08}. Remarkably, both predict a $T^2$- and
$B^2$-dependence of the linewidth in the LFL regime which is indeed found experimentally, albeit at fields and below
temperatures close to the extrapolated $T^*$ line (Fig.~\ref{width}b,c, Fig.~\ref{diag_spectra}, left frame).

\indent In summary, we have studied HF-ESR in \YRS\ in a large $T-B$-parameter domain extending from the breakdown of the heavy fermion state at
large $T$ and large fields until the crossover to the LFL state at lower temperatures. By comparing  the $T$- and $B$-dependences of the Sommerfeld
coefficient \cite{Trovarelli00,Gegenwart06}, the $^{29}$Si-NMR Knight shift and $1/T_1$ rate \cite{Ishida02} with those of the ESR $g$-factor
and the linewidth $\Delta B$, the signatures of this crossover have been identified in the ESR measurables. The striking similarity of these
dependences strongly suggests that the ESR response is due to a resonance excitation of the quasiparticles in the Kondo lattice, i.e. the ''heavy
electron spin resonance''. As such it probes thus directly the evolution of the Kondo state and the occurrence of different electronic regimes in
\YRS. This conjecture qualitatively explains the magnetic field dependence of: (i) - the $g$-factor in terms of a gradual change from a
$T$-independent towards a $\ln T$-dependent $g$-factor; (ii) - the line broadening, in particular, a crossover to a $T^2$-variation of $\Delta B$ at
low $T$ in strong fields.

\indent This work was supported by the DFG through SFB 463 and the Research Unit 960. The work of AZ was supported in part by the DFG grant 436 UKR
17/21/06. JS acknowledges helpful discussions with E. Abrahams.

\end{document}